# A hierarchical Algorithm to Solve the Shortest Path Problem in Valued Graphs

Michel Koskas[*]

**Abstract**

This paper details a new algorithm to solve the shortest path problem in valued graphs. Its complexity is $O(D \log v)$ where $D$ is the graph diameter and $v$ its number of vertices. This complexity has to be compared to the one of the Dijkstra's algorithm, which is $O(e \log v)$ where $e$ is the number of edges of the graph. This new algorithm lies on a hierarchical representation of the graph, using radix trees. The performances of this algorithm show a major improvement over the ones of the algorithms known up to now.

## 1 Introduction

We present in this paper a new algorithm[1] allowing one to find the shortest path between two edges. This problem is one of the oldest and more studied of the graph area. A lot of papers have been written on this subject (see [1], [5], [6], [14] or [13] for instance). Some algorithms span graphs with trees or use spanning trees (see [7] for instance), and the multicast problem is also very studied (see for instance [2], [3], [9], [10], [11], [12], [15]).

One can imagine a graph as a set of cities (or Internet machines) (the vertices) and roads (the edges). The length of the road between two cities may be seen as the cost function (or *valuation*) of the edge. A path consists in a sequence of vertices related one to the next by a road and the path cost is the sum of the lengths of the used edges. The path problem is to try to find the shortest path between two vertices.

The valuation of an edge is usually a non negative real number but one can imagine that it could (at least theoretically) be negative. The Dijkstra's algorithm does not work in this case. The present algorithm may work if slightly modified and if the graph does not contain any cycle of negative length (in this last case, the shortest path problem looses its meaning because for any couple of vertices $v$ and $v'$ such that there exists

[*]koskas@laria.u-picardie.fr

[1]patent pending

a path between $v$ and $v'$ that intersects this cycle, one can find a path of lower valuation (by running through the cycle as many times as wanted)).

Now, we suppose that the valuations of the edges are all non negative real numbers.

What if we approximate the valuations with rational numbers precisely enough to assure that two distinct numbers are approximated with distinct rationals ?

In this case, it is equivalent to use the real valuations or their approximations for looking for one path. Then by multiplying all the valuations by the lcm (lowest common multiple) of the denominators of the valuations, we have to solve the shortest path problem by using integers valuations.

This is why in this paper we shall discuss only the cases of a graph not valued (all the edges are valued by 1) or the case in which the edges are valued with non positive integers.

The best known algorithm up to now to solve the shortest path problem is the Dijkstra's algorithm. It runs in time $O(V \ln E)$ where $V$ is the number of vertices of the graph and $E$ the number of edges (see [4], [8] or [16]).

We present here an algorithm which improves dramatically the time needed to solve this problem. Its complexity is $O(l \ln V)$ where $l$ is the average shortest path length and $V$ is the number of vertices of the graph.

In the first section, we give a new description of a graph. In the second section, we explain how this description may be used and we develop an algorithm to find $k$-extreme paths between two vertices, in the case the graph is not valued.

In the third section, we give a more efficient version of the algorithm, which lies on a hierarchization of the graph. The computation time is majored by $C(l \ln v)$ where $C$ is a constant, $v$ the number of vertices of the graph and $l$ the average path-length.

In the fourth section, we generalize the algorithm in the case of discretely valued graphs.

We conclude in the fifth section.

## 2 Graphs: a new description

We first recall usual definitions useful to someone dealing with graphs. Then we give a new definition of graphs and vertices, and give an example of a graph with the usual and this new definitions.

### 2.1 Usual definitions

An unvalued graph is a couple $(\mathcal{V}, \mathcal{E})$ where $\mathcal{V}$ is a finite set, the *vertices*, and $\mathcal{E}$ a set of couples of vertices. The elements of $\mathcal{E}$ are called *edges*. The first element of an edge is called *origin* of the edge. Its second element is called *extremity*.

When $(v_1, v_2) \in \mathcal{E} \Rightarrow (v_2, v_1) \in \mathcal{E}$, the graph is said to be *undirected* and $\mathcal{E}$ may be considered as a set of pairs of vertices.

A path between two vertices $v_1$ and $v_2$ is a sequence $u_0, \ldots u_k$ of elements of $\mathcal{V}$ such that $u_0 = v_1$, $u_k = v_2$ and for all $i$ in $[\![0, k-1]\!]$, $(u_i, u_{i+1}) \in \mathcal{E}$. In this case, $k$ is said to be the *length* of the path.

The path-problem is to find the shortest path between two vertices of the graph, *i.e.* in this case the path, or the paths, using as few vertices as possible.

A valued graph is a couple $(\mathcal{V}, \mathcal{E})$ where $\mathcal{V}$ is a finite set, the vertices, and $\mathcal{E}$ a set of triples $(v, v', x)$ where $v, v' \in \mathcal{V}$ and $x \in \mathbb{R}$, the valued edges (or when not ambiguous, the edges). Sometimes, $x$ is named the *cost*, or even the *length*, of the edge.

A path between two vertices $v_1$ and $v_2$ is a sequence $u_0, \ldots u_k$ of vertices such that $u_0 = v_1$, $u_k = v_2$ and for all $i$ in $[\![0, k-1]\!]$, $\exists x_i \in \mathbb{R}$, $(u_i, u_{i+1}, x_i) \in \mathcal{E}$. The length of the path is in this case $\sum_{i=0}^{k-1} x_i$.

All occurs as if $x_i$ were the length of the edge and $\sum_{i=0}^{k-1} x_i$ were the length of the path.

The path problem is to find the shortest path between two vertices.

## 2.2 A new definition

Given an unvalued graph $G = (\mathcal{V}, \mathcal{E})$, let us consider the followings maps:

Let $V = \#(\mathcal{V})$, and let us denote the vertices of $\mathcal{V}$ by $v_1, v_2, \ldots, v_V$.

Now let us map $\mathcal{V}$ to $(\mathbb{Z}/2\mathbb{Z})^V$ by $v_i \mapsto e_i$, $(e_i)_{1 \leq i \leq V}$ being the canonical basis of $(\mathbb{Z}/2\mathbb{Z})^V$.

So from now on, we consider that the vertices are orthogonal vectors of $(\mathbb{Z}/2\mathbb{Z})^V$ ; hence we shall identify $\mathcal{V}$ and $(\mathbb{Z}/2\mathbb{Z})^V$.

Let $\mathcal{E}$ be the dual space of $\mathcal{V}$.

Now we consider that the edges are elements of $\mathcal{E}$.

We distinguish two sets of edges: the outgoing and the incoming edges.

Let us denote by $O$ the set of outgoing edges which is designed as follows. For all $v \in \mathcal{V}$ one has a linear form $o_v \in \mathcal{E}$ such that $\forall v' \in \mathcal{V}, o_v(v') = \begin{cases} 1 & \text{if } (v, v') \in \mathcal{E} \\ 0 & \text{otherwise} \end{cases}$.

The set $I$ of incoming edges, now: for all $v \in \mathcal{V}$, one has a linear form $i_v \in \mathcal{E}$ such that $\forall v' \in \mathcal{V}, i_v(v') = \begin{cases} 1 & \text{if } (v', v) \in \mathcal{E} \\ 0 & \text{otherwise} \end{cases}$.

**Remark** $(\mathcal{V}, O, I)$ describe the graph as well as $(\mathcal{V}, \mathcal{E})$. Is also true for $(\mathcal{V}, O)$ or for $(\mathcal{V}, I)$.

**Example** The following figure shows a graph and its representation.

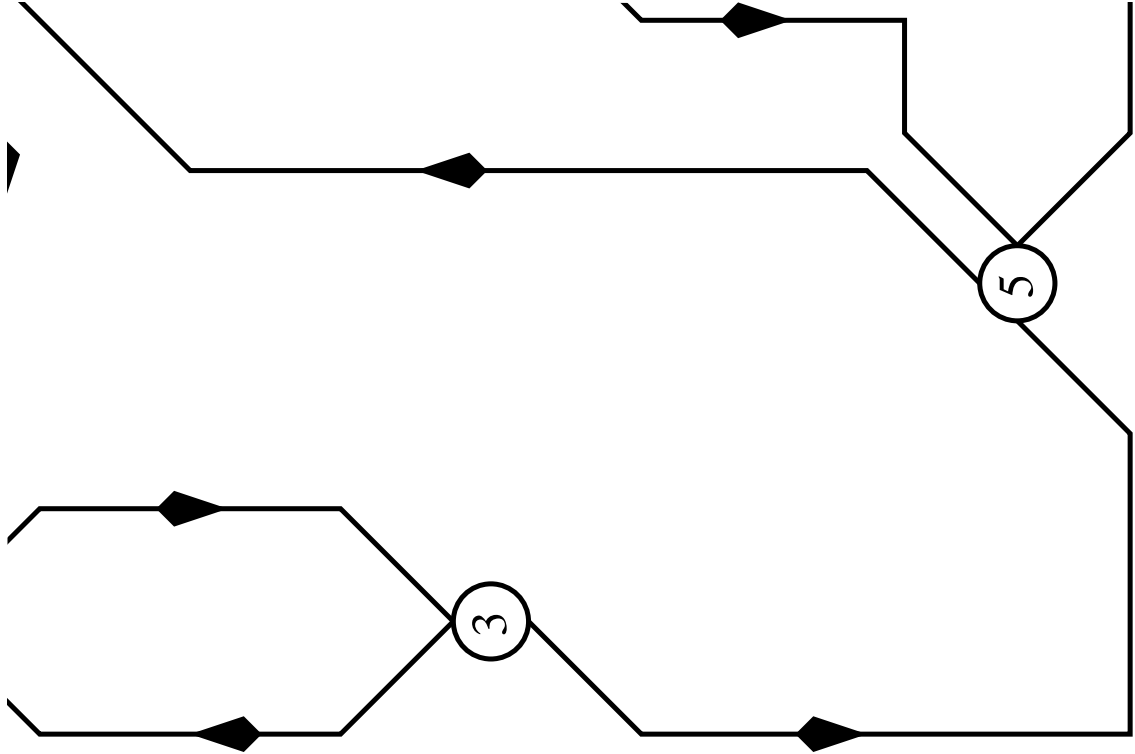

Its adjacency matrix is:

$$Adj = \begin{pmatrix} 0 & 1 & 1 & 0 & 0 \\ 1 & 0 & 0 & 0 & 1 \\ 1 & 0 & 0 & 0 & 0 \\ 0 & 0 & 0 & 0 & 1 \\ 0 & 1 & 0 & 1 & 0 \end{pmatrix}, \text{ and } Adj^t = \begin{pmatrix} 0 & 1 & 1 & 0 & 0 \\ 1 & 0 & 0 & 0 & 1 \\ 1 & 0 & 0 & 0 & 0 \\ 0 & 0 & 0 & 0 & 1 \\ 0 & 1 & 0 & 1 & 0 \end{pmatrix}.$$

The edges are:

$$O = \{ \begin{pmatrix} 0 \\ 1 \\ 1 \\ 0 \\ 0 \end{pmatrix}, \begin{pmatrix} 1 \\ 0 \\ 0 \\ 1 \\ 0 \end{pmatrix}, \begin{pmatrix} 1 \\ 0 \\ 0 \\ 0 \\ 1 \end{pmatrix}, \begin{pmatrix} 0 \\ 0 \\ 0 \\ 0 \\ 1 \end{pmatrix}, \begin{pmatrix} 0 \\ 1 \\ 0 \\ 1 \\ 0 \end{pmatrix} \}$$

and

$$I = \{ \begin{pmatrix} 0 \\ 1 \\ 1 \\ 0 \\ 0 \end{pmatrix}, \begin{pmatrix} 1 \\ 0 \\ 0 \\ 0 \\ 1 \end{pmatrix}, \begin{pmatrix} 1 \\ 0 \\ 0 \\ 0 \\ 0 \end{pmatrix}, \begin{pmatrix} 0 \\ 1 \\ 0 \\ 0 \\ 1 \end{pmatrix}, \begin{pmatrix} 0 \\ 0 \\ 1 \\ 1 \\ 0 \end{pmatrix} \}.$$

## 2.3 Mixing edges

Let us now defines a few laws among edges and among vertices.

$$\forall o, o' \in \mathcal{E}, \, o \vee o' = (o_i \vee o'_i)_{1 \leq i \leq H}$$

This law is a logical "or" for the edges: for all vertices $v$ and $v'$, $(o_v \vee o_{v'})_i = 1$ if and only if there exists an edge $(v, e_i)$ or an edge $(v', e_i)$ (we recall that $(e_i)_{1 \leq i \leq V}$ is the canonical basis of $\mathcal{V}$).

Let us consider furthermore the following law:

$$\forall o, o' \in \mathcal{E}, \, o \wedge o' = (o_i \wedge o'_i)_{1 \leq i \leq H}.$$

This law is a logical "and" for the edges: for all vertices $v$ and $v'$, $(o_v \wedge o_{v'})_i = 1$ if and only if there exists an edge $(v, e_i)$ *and* an edge $(v', e_i)$.

Let us denote by $v$ the operation mapping any element $e$ of $\mathcal{E}$ to the set $v(e) = \{w \in \mathcal{V}, e(w) = 1\}$.

We define now an incrementation and a decrementation over edges:

$$\forall e \in \mathcal{E}, \forall k \in \mathbb{N}, \, e + k = \begin{cases} e & \text{if } k = 0, \\ \vee_{e' \in v(e)} o_{e'} & \text{if } k = 1, \\ (e + (k-1)) + 1 & if k \geq 2 \end{cases}$$

and

$$\forall e \in \mathcal{E}, \forall k \in \mathbb{N}, \, v - k = \begin{cases} e & \text{if } k = 0, \\ \vee_{e' \in v(e)} i_{e'} & \text{if } k = 1, \\ (e - (k-1)) - 1 & if k \geq 2 \end{cases}.$$

We similarly define an incrementation and a decrementation for the elements of $\mathcal{V}$:

$$\forall v \in \mathcal{V}, \forall k \in \mathbb{N}, \, v + k = \begin{cases} v^* & \text{if } k = 0, \\ o_v & \text{if } k = 1, \\ o_v + (k-1) & \text{if } k \geq 2. \end{cases}$$
$$(v^* \text{ being the dual of } v)$$

**Lemma 1** *For any $v, v' \in \mathcal{V}$, for any $k \geq 0$, there exists a path of length $k$ between $v$ and $v'$ if and only if $(v + k)(v') = 1$. Similarly, there exists a path of length $k$ between $v$ and $v'$ if and only if $(v' - k)(v) = 1$.*

**Proof:** if $k = 1$, the result holds by definition of $v + 1$. If $k > 1$, there exists a path of length $k$ between $v$ and $v'$ if and only if there exists a vertex $v''$ such that there exists a path of length $k-1$ between $v$ and $v''$ and a path of length 1 between $v''$ and $v'$. By the definition of $v + (k-1)$ we have that $v + (k-1)(v'') = 1$ and $v'' + 1(v') = 1$. Thus, $v + k(v') = 1$. Conversely, if $v + k(v') = 1$ there exists a vertex $v''$ such that $v + (k-1)(v'') = 1$ and $v'' + 1(v') = 1$. By induction, there exists a path of length $k-1$ between $v$ and $v''$ and by definition a path of length 1 between $v''$ and $v'$.

**Theorem 1** *Let $v$ and $v'$ be two vertices and $k$ an integer. There exists a path of length $k$ between $v$ and $v'$ if and only if for any $j$ such that $0 \leq j \leq k$, $v((o_v + j) \wedge (i_{v'} - (k-j))) \neq \emptyset$.*

**Proof:** Let $v$ and $v'$ be two vertices such that there exists a path of length $k$ between $v$ and $v'$. Let us denote such a path $v_0 = v$, $v_1$, ... $v_k = v'$. Then by definition, for any $0 \leq j \leq k-1$, $v_j + 1(v_{j+1}) = 1$ and $v_{j+1} - 1(v_j) = 1$. By induction, for any $j$ such that $0 \leq j \leq k$, $v((o_v + j) \wedge (i_{v'} - (k-j))) \neq \emptyset$. Indeed, this set contains at least $v_j$.

**Theorem 2** *Let $v$ and $v'$ be two vertices and $k$ an integer. Then if for any $j \leq k$, $v((o_v + j) \wedge (i_{v'} - (k-j))) \neq \emptyset$ then $\forall j' \in [\![0, k]\!]$, $v((o_v + j').(i_{v'} - (k-j'))) \neq \emptyset$.*

**Proof:** If for any $0 \leq j \leq k$, $v((o_v + j) \wedge (i_{v'} - (k-j))) \neq \emptyset$, then it is true for any $0 \leq j' \leq k$. Indeed, let $w \in v((o_v + j) \wedge (i_{v'} - (k-j)))$. Then there exists a path of length $j$ between $v$ and $w$ and a path of length $k-j$ between $w$ and $v'$. So there exists a path of length $k$ between $v$ and $v'$.

We are now ready to develop the underlying algorithms.

## 3 Finding Paths in Unvalued Graphs

In this section we will detail an algorithm which computes the shortest path between two vertices of a graph when its edges are not valued, *i.e.* when the cost-function is the number of used vertices.

Let $G = (\mathcal{V}, O, I)$ be a graph. Let us denote as preceedingly $V = \#(\mathcal{V})$.

Let $v_1$ and $v_2$ be two vertices. The following algorithm computes the shortest path between $v_1$ and $v_2$. The computation time is majored by $lV$ where $l$ is the path length and $V$ is the number of vertices of the graph.

```
ShortestPath1(G, v[1], v[2])
Begin
if (v[1] = v[2]) PrintPath(v[1], 1)
EndIf
l = 0
o[l] = v[1]
Repeat
   l = l + 1
   o[l] = o[l-1] + 1
Until ((v[1] + l)(v[2]) = 1) Or (o[l-1] = o[l])
if (o[l - 1] = o[l])
   print(``The graph is not strongly connected and'')
   print(``there is no path between v1 and v2'')
   Return
EndIf
```

```
LengthPath = l
i[0] = v[2]
while (l > 0) Do
   o[l] ∧= i[LengthPath - l]
   i[LengthPath - l + 1] = i[LengthPath - l] - 1
   l--
EndWhile
ReadPathes(LengthPath, o)
End
```

The function ReadPathes is a simple deep-first tree reading:

```
ReadPathes(LengthPath, o[])
Begin
deepness = 0
For i = 0 to LengthPath p[i] = 0 EndFor
DeepFirst(p, deepness, LengthPath, o)
End

DeepFirst(p[], deepness, LengthPath, o[])
Begin
if (deepness = LengthPath)
   PrintPath(p, LengthPath)
   Return
EndIf
Foreach h in o(o[l])
   p[deepness] = h
   Recall = o[deepness + 1]
o[deepness + 1] ∧= o[h]
ReadPath(deepness + 1, LengthPath, o)
o[deepness + 1] = Recall
EndForEach
End

PrintPath(p[], LengthPath)
Begin
for i = 1 to LengthPath - 1
   printf(p[i], ``->'')
printf(p[LengthPath])
End
```

# 4 finding Paths in Valued Graphs

Let $G = (\mathcal{V}, \mathcal{E})$ be a graph. One can build the same sets of vectors and linear forms than done, as if the edges had a valuation of 1 and store the cost function, *i.e.* the valuations of the edges. Of course, this cost function may be extended to paths (by summing the costs of the couples of vertices constituting the path).

A valued graph is then described by is vertices, its outgoing (or incoming) edges and its cost function.

So we can talk about graphs as a quadruple $G = (\mathcal{V}, O, I, c)$ where $\mathcal{V}$ is the set of vertices, $O$ and $I$ are the sets of outgoing and incoming edges, and $c$ the cost function which maps a couple of vertices to a non negative real (or, as we saw in the introduction, to an integer) number.

Let $G = (\mathcal{V}, O, I, c)$ be a valued graph, in which the edges have integer weights.

Given a path $p$, let us consider the two cost functions of $p$: $ne(p)$ is the number of hedges occurring in $p$ and $c(p)$ is the total weight of $p$.

As the edges are valued with integers, it is not useful to try to find a path using more than $c(p)$ edges.

Consequently we can propose the following algorithm:

```
ShortestPath2(G, v[1], v[2])
Begin
Find among the paths p(v[1],v[2]) of minimal ne(p), the one
  (or the ones) minimizing c

Set the extra-vertices number (evn) to evn = c(p) - ne(p)
j = 1
While (j <= evn) Do
   find all paths p' of length ne(p) + j minimizing c.
   If c(p') < c(p),
      p = p'
      evn = c(p) - ne(p)
   EndIf
EndWhile
End
```

This algorithm gives the shortest path in a graph valued by integers.

# 5 Graph Trees : a Hierarchization

In this section, we explain how one can use the preceeding algorithms to find in linear time the shortest path between two vertices. We still work in $G = (\mathcal{V}, O, I)$ and we suppose that the edges of $G$ are not valued.

The idea of the algorithm is to try to work on a graph much littler than $G$. To achieve this goal, we will use equivalence relations and find paths among the matching equivalence classes in $G/\sim$. Then the path between the vertices of $G$ can be found among the rises of the path found in $G/\sim$.

## 5.1 A first reduction

Let $G = (\mathcal{V}, \mathcal{E})$ be a graph and let $\sim$ be an equivalence relation among the vertices of $G$. Let us consider the graph $G' = (\mathcal{E}, \mathcal{E}')$ whose vertices are the equivalence classes of $G/\sim$. Given two vertices $v_1$ and $v_2$ of $G'$, the edge $(v_1, v_2) \in \mathcal{E}'$ if and only if there exists $w_1 \in v_1 \subset \mathcal{V}$ and $w_2 \in v_2 \subset \mathcal{V}$ such that $(w_1, w_2) \in \mathcal{E}$.

We shall call such a graph $G'$ a *thickening* of $G$ and we shall say that $G$ is a refinement of $G'$.

**Theorem 3** *Let $v_1$ and $v_2$ be two vertices of $\mathcal{V}$. If there exists a path driving from $v_1$ to $v_2$ then there exists a path in $G'$ driving from $\overline{v_1}$ to $\overline{v_2}$ where $\overline{v}$ is the equivalence class of $v \in \mathcal{V}$.*

**proof:** indeed, let $w_0 = v_1, w_1, \ldots, w_k = v_2$ be a path between $v_1$ and $v_2$ and let $w'_i$ be the equivalence class of $w_i$ for $0 \le i \le k$. Then $w'_0, w'_1, \ldots, w'_k$ is a path of length $k$ between $\overline{v_1}$ to $\overline{v_2}$.

**Remark 1** *The reciprocal proposition is false. For instance if a graph is not strongly connected and any couple of vertices are equivalent with the $\sim$ relation, there exists a path between any couple of vertices of $G'$ which is not true for any couple of vertices of $G$.*

**Theorem 4** *Let $v_1$ and $v_2$ be two vertices of $\mathcal{V}$. If there exists a path of length $k$ driving from $v_1$ to $v_2$ then there exists a path of length $k$ in $G'$ driving from $\overline{v_1}$ to $\overline{v_2}$ where $\overline{v}$ is the equivalence class of $v \in \mathcal{V}$.*

**Remark 2** *The reciprocal proposition is false. It may so happen that there exists a path of length $k$ between $\overline{v_1}$ and $\overline{v_2}$ and that there exists a longer path or no path at all between $v_1$ and $v_2$.*

**Remark 3** *Let $v_1$ and $v_2$ be two vertices of $\mathcal{V}$ and $\overline{v_1}$ and $\overline{v_2}$ their equivalences classes in $\mathcal{E}$. Let $k$ be the length of the shortest path (i.e. using as few edges as possible) between $v_1$ and $v_2$, and $k'$ be the length of the shortest path between $\overline{v_1}$ and $\overline{v_2}$. Then $k \ge k'$.*

**Definition 1** *Let $G = (\mathcal{V}, \mathcal{E})$ be a graph and let $G' = (\mathcal{E}, \mathcal{E}')$ be a thickening of $G$. Let $v_1$ and $v_2$ be two elements of $\mathcal{V}$ and let us denote as usual $\overline{v_1}$ and $\overline{v_2}$ their equivalences classes in $\mathcal{E}$. Let $p = (u_0, \ldots, u_k)$ be a path between $v_1$ and $v_2$ in $G$, and let $p' = (u'_0, \ldots, u'_k)$ a path between $\overline{v_1}$ and $\overline{v_2}$ in $G'$. We shall say that $p$ is a refinement of $p'$, or, equivalently, that $p'$ is a thickening of $p$ is $\forall 0 \le j \le k$, $u_j \in \overline{u_j}$.*

## 5.2 Computing paths refinements

In this section we detail how using equivalence relations may be useful to compute efficiently the shortest path between two vertices.

**Definition 2** *Let $\overline{v}$ be a vertex of $\mathcal{E}$. The dumb refinement of $\overline{v}$ is the edge (of $\mathcal{V}$) $d_{\overline{v}}$ such that $d_{\overline{v}}(v') = 1 \Leftrightarrow v' \in \overline{v}$. Let $\overline{e}$ be an edge of $\mathcal{E}'$. The dumb refinement of $\overline{e}$ is the edge $\vee_{e(v)=1} e_v$. If $\overline{e}$ is an outgoing vertex, we shall speak of a dumb outgoing refinement, and if $\overline{e}$ is an incoming edge, we shall speak of an incoming dumb refinement.*

Let $v_1$ and $v_2$ be two elements of $\mathcal{V}$ and let $\overline{v_1}$ and $\overline{v_2}$ be their equivalence classes.

Let $\overline{p}$ be a path between $\overline{v_1}$ and $\overline{v_2}$. The following algorithm computes the refinements of $\overline{p}$ whenever exists.

Let us denote $\overline{p} = (\overline{u}_0, \ldots, \overline{u}_k)$, and let $e'_i$ be the dumb refinement of the edge $(u_i, u_{i+1}$. Then one has simply to compute $e'_i \wedge (v_1 + i) \wedge (v_2 - (k-1-i))$. If one of these edges is the null linear form, then there does not exist such a refinement. Otherwise, the edges may be expanded in paths as explained in a preceeding section.

So the algorithm is:
```
Refine(p :  path, v1, v2 :  vertices,
    w1, w2 :  EquivalencesClasses, PathLength :  Integer)
Begin
Forward[], BackWard[], Res[] :  edges
Forward[0] = Outgoing(v1)
BackWard[PathLength] = Incoming(v2)
For i = 1 to PathLength - 1 Do
   Forward[i] = Forward[i - 1] + 1
   Res[i] = p[i] ∧ Forward[i]
EndFor
For i = PathLength - 1 Downto 0 Do
   BackWard[i] = BackWard[i + 1] - 1
   Res[i] ∧ = Bacward[i]
EndFor
If Nul(Res[])
   Print(``No Possible Refinement'')
Else
   PrintPathes(Res[], PathLength)
EndIf
End
```
And now, we are ready to produce an algorithm using the equivalences classes.

$G$ is a graph, $G'$ is a thickening of $G$. We have two vertices $v_1$ and $v_2$ and we are looking for the shortest path between $v_1$ and $v_2$. As usually, we denote by $\overline{v_1}$ and $\overline{v_2}$ the equivalences classes of $v_1$ and $v_2$.

```
ShorterPath3(v1, v2)
Begin
PathFound = False
p[] : path
PathLength = 0
While (Not PathFound)
   FindPathesOfFixedLength(v1̄, v2̄, PathLength, p)
   ForEach path in p
      If (Refine(path))
         printf(``Path Found'')
      EndIf
   EndForEach
EndWhile
End
```

Finally, we can use not only one thickening but many thickenings of thickenings. This means to say that we can build a sequence of graphs $g_0 = G$, $G_1$, ... $G_k$ and find a path in $G_k$ that is to be refined in $G_{k-1}$ and this refined path has to be refined in $G_{k-2}$ and so on...

Usually when a path can not be refined, it may be seen immediately, because it is very uncommon that two vertices of the same equivalence class have outgoing vertices to the same vertex.

Hence, the complexity of this algorithm is linear in the length of the path.

## 5.3 Example

Let us consider the following graph $G_0$:

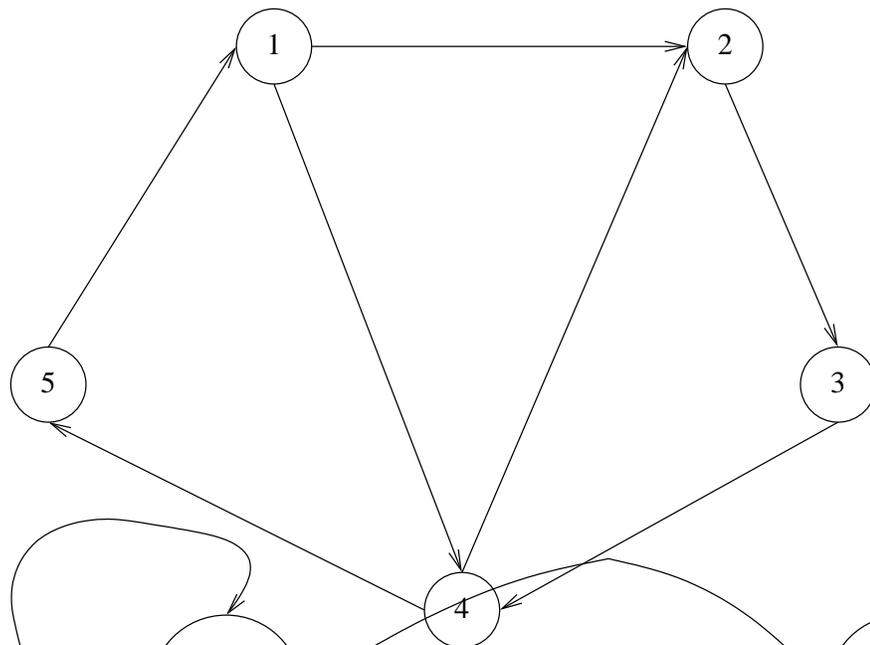

By grouping the vertices 1 and 2 in the one hand and 3 and 4 in the other hand, on obtains the graph $G_1$:

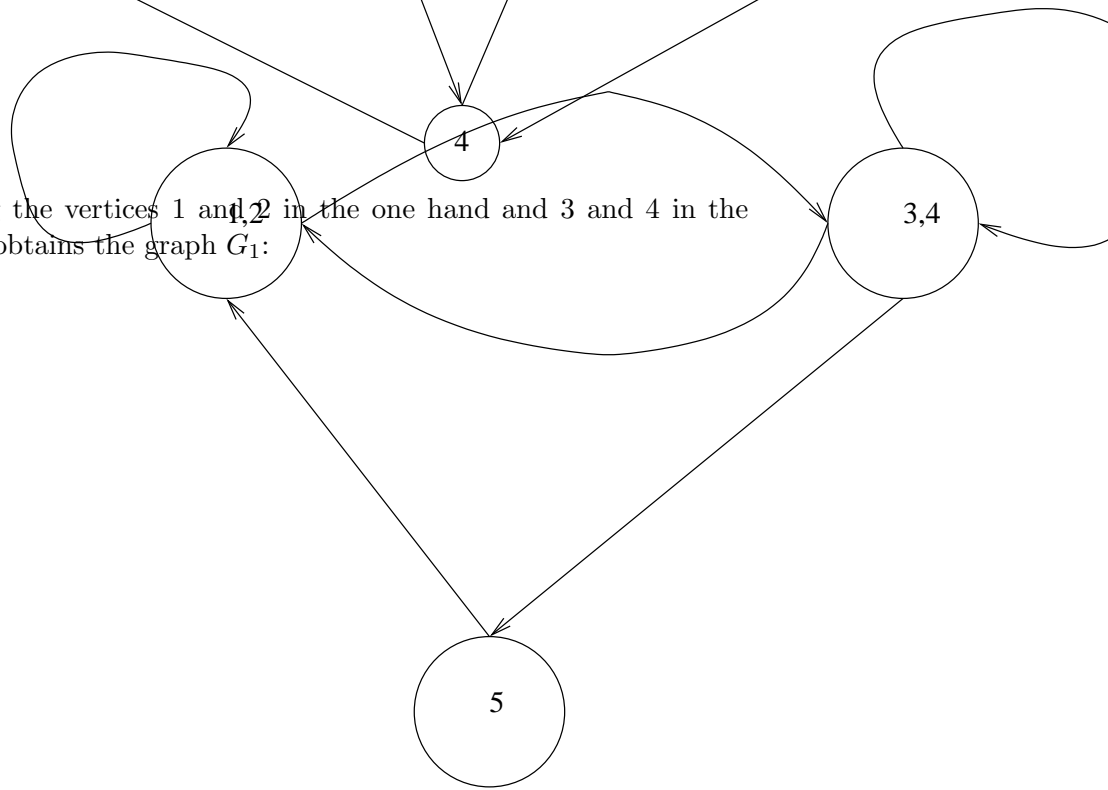

By grouping again the vertices two by two, one obtains the graph $G_2$:

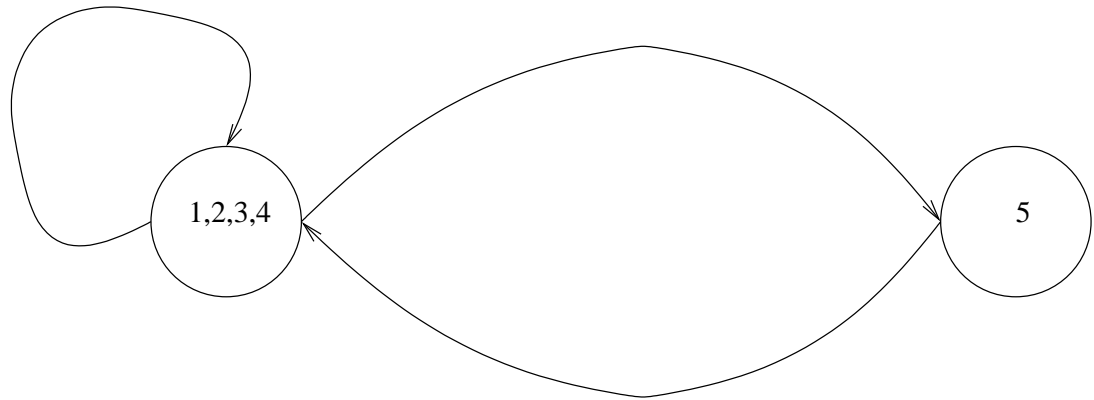

By grouping a last time the vertices two by two, one obtains the graph $G_3$:

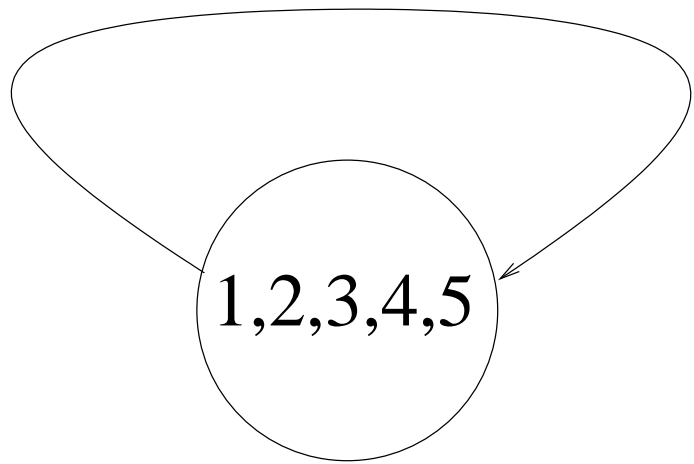

In $G0$, the outgoing and incoming edges are:

$O_0 = \{o_{v_1} = \begin{pmatrix} 0 \\ 1 \\ 0 \\ 1 \\ 0 \end{pmatrix}, o_{v_2} = \begin{pmatrix} 0 \\ 0 \\ 1 \\ 1 \\ 0 \end{pmatrix}, o_{v_3} = \begin{pmatrix} 0 \\ 0 \\ 0 \\ 1 \\ 0 \end{pmatrix}, o_{v_4} = \begin{pmatrix} 0 \\ 0 \\ 0 \\ 0 \\ 1 \end{pmatrix}, o_{v_5} = \begin{pmatrix} 1 \\ 0 \\ 0 \\ 0 \\ 0 \end{pmatrix}$ and

$$I_0 = \{i_{v_1} = \begin{pmatrix} 0 \\ 0 \\ 0 \\ 0 \\ 1 \end{pmatrix}, i_{v_2} = \begin{pmatrix} 1 \\ 0 \\ 0 \\ 0 \\ 0 \end{pmatrix}, i_{v_3} = \begin{pmatrix} 0 \\ 1 \\ 0 \\ 0 \\ 0 \end{pmatrix}, i_{v_4} = \begin{pmatrix} 1 \\ 1 \\ 1 \\ 0 \\ 0 \end{pmatrix}, i_{v_5} = \begin{pmatrix} 0 \\ 0 \\ 0 \\ 1 \\ 0 \end{pmatrix}.$$

In $G_1$, the outgoing and incoming vertices are:
$$O_1 = \{o_{v_1,v_2} = \begin{pmatrix} 1 \\ 1 \\ 0 \end{pmatrix}, o_{v_3,v_4} = \begin{pmatrix} 0 \\ 1 \\ 1 \end{pmatrix}, o_{v_5} = \begin{pmatrix} 1 \\ 0 \\ 0 \end{pmatrix} \text{ and}$$

$$I_1 = \{i_{v_1,v_2} = \begin{pmatrix} 1 \\ 0 \\ 1 \end{pmatrix}, i_{v_3,v_4} = \begin{pmatrix} 1 \\ 1 \\ 0 \end{pmatrix}, i_{v_5} = \begin{pmatrix} 0 \\ 1 \\ 0 \end{pmatrix}.$$

Finally, in $G_2$, the outgoing and incoming vertices are:
$$O_2 = \{o_{v_1,v_2,v_3,v_4} = \begin{pmatrix} 1 \\ 1 \end{pmatrix}, o_{v_5} = \begin{pmatrix} 1 \\ 0 \end{pmatrix} \text{ and}$$

$$I_2 = \{i_{v_1,v_2,v_3,v_4} = \begin{pmatrix} 1 \\ 1 \end{pmatrix}, i_{v_5} = \begin{pmatrix} 1 \\ 0 \end{pmatrix}.$$

In $G_3$, the outgoing and incoming vertices are:
$O_3 = \{o_{v_1,v_2,v_3,v_4,v_5} = (\,1\,)\}$ and
$I_3 = \{i_{v_1,v_2,v_3,v_4,v_5} = (\,1\,)\}$.

We will denote indifferently the class of the edge $i$ by $\bar{i}$ (context gives the thickening we are talking about) or for instance by $(1,2)$ for the class of 1 (or 2) in the first thickening, $((1,2),(3,4))$ for the class of 1 (or 2 or 3 or 4) in the second thickening, and so on...

Let us compute the shortest path between $v_2$ and $v_5$ for instance.

Let us look for a path of length 1 between these two vertices.

At level 3, there exists a path of length 1 between $\bar{2}$ and $\bar{5}$, which is $\bar{2} \to \bar{5}$.

At level 2, we "and" $\bar{2}+1$ and $\bar{5}$. This is $o_{\bar{2}+1} \wedge \bar{5} = \begin{pmatrix} 0 \\ 1 \end{pmatrix}$ which is not null.

Then we "and" $\bar{5}-1$ and $\bar{2}$, which gives $i_{\bar{5}}-1 \wedge \bar{2} = \begin{pmatrix} 1 \\ 0 \end{pmatrix}$ which is not null.

So at level 2, we have a path of length 1 between $\bar{2}$ and $\bar{5}$, which is (of course) $\bar{2} \to \bar{5}$.

At level 1, we have then to intersect $\bar{2}+1$ and $\bar{5}$. This gives: $\bar{2}+1 \wedge \bar{5} = \begin{pmatrix} 1 \\ 1 \\ 0 \end{pmatrix} \wedge \begin{pmatrix} 0 \\ 0 \\ 1 \end{pmatrix} = \begin{pmatrix} 0 \\ 0 \\ 0 \end{pmatrix}.$

So we conclude here that there is no path of length 1 at level 1 between $\overline{2}$ and $\overline{5}$, and neither at level 0 between $v_2$ and $v_5$.

Let us look for a path of length 2 between $v_2$ and $v_5$.

At level 3, there is a path of length 2 between $\overline{2}$ and $\overline{5}$. This path is $(((1,2),(3,4)),5) \to (((1,2),(3,4)),5) \to (((1,2),(3,4)),5)$.

At level 2, this drives to compute $\overline{2} + 1 \wedge d(v_{1,2,3,4,5}) \wedge overline{5} - 1$. So we have to compute $\begin{pmatrix} 1 \\ 1 \end{pmatrix} \wedge \begin{pmatrix} 1 \\ 1 \end{pmatrix} \wedge \begin{pmatrix} 1 \\ 0 \end{pmatrix} = \begin{pmatrix} 1 \\ 0 \end{pmatrix}$.

So at level 2, there is a path of length 2 between $\overline{2}$ and $\overline{5}$, which is $\overline{2} \to \overline{2} \to \overline{5}$.

At level 1, we have to compute $\overline{2} + 1 \wedge d(\overline{2}) \wedge \overline{5} - 1$. This gives: $\begin{pmatrix} 1 \\ 1 \\ 0 \end{pmatrix} \wedge \begin{pmatrix} 1 \\ 1 \\ 0 \end{pmatrix} \wedge \begin{pmatrix} 0 \\ 1 \\ 0 \end{pmatrix} = \begin{pmatrix} 0 \\ 1 \\ 0 \end{pmatrix}$.

So at level 1, there exists a path between $\overline{2}$ and $\overline{5}$, which is $\overline{2} \to \overline{3} \to \overline{5}$.

Let us look for a refinement of this path at level 0:

Let us compute $v_2 + 1 \wedge d(\overline{3}) \wedge v_5 - 1 = \begin{pmatrix} 0 \\ 0 \\ 1 \\ 1 \\ 0 \end{pmatrix} \wedge \begin{pmatrix} 0 \\ 0 \\ 1 \\ 1 \\ 0 \end{pmatrix} \wedge \begin{pmatrix} 0 \\ 0 \\ 0 \\ 1 \\ 0 \end{pmatrix} = \begin{pmatrix} 0 \\ 0 \\ 0 \\ 1 \\ 0 \end{pmatrix}$.

So there is a path of length 2 between $v_2$ and $v_5$ which is $v_2 \to v_4 \to v_5$.

We fully detail a refinement now: The refinements are computed as explained before: let us refine for instance the path

$$((1,2),(3,4)) \to ((1,2),(3,4)) \to ((5))$$

from $G_2$ in $G_1$.

$(1,2) + 0 = \begin{pmatrix} 1 \\ 0 \\ 0 \end{pmatrix}$

$(1,2) + 1 = \begin{pmatrix} 1 \\ 1 \\ 0 \end{pmatrix}$

and $(1,2) + 2 = \begin{pmatrix} 1 \\ 1 \\ 0 \end{pmatrix}$.

Then $e_2 = (1,2) + 2 \wedge (5) - 0 = \begin{pmatrix} 0 \\ 0 \\ 1 \end{pmatrix}$,

Then $e_1 = (1,2) + 1 \wedge (5) - 1 \wedge dumb(((1,2),(3,4))) = \begin{pmatrix} 0 \\ 1 \\ 0 \end{pmatrix}$,

and $e_0 = (1,2) + 0 \wedge (5) - 2 \wedge dumb(((1,2),(3,4))) = \begin{pmatrix} 1 \\ 0 \\ 0 \end{pmatrix}$,

Then there is one refinement of $((1,2),(3,4)) \to ((1,2),(3,4)) \to ((5))$ from $G_2$ to $G_1$ which is $(1,2) \to (3,4) \to (5)$.

We can similarly compute the refinement of this path from $G_1$ to $G_0$ :

$$f_1 = v_2 + 1 \wedge dumb(3,4) = \begin{pmatrix} 0 \\ 0 \\ 1 \\ 0 \\ 0 \end{pmatrix},$$

$$v_2 + 2 \wedge v_5 = \begin{pmatrix} 0 \\ 0 \\ 0 \\ 0 \\ 1 \end{pmatrix},$$

and then $f_1 \wedge v_5 - 1 = \begin{pmatrix} 0 \\ 0 \\ 1 \\ 0 \\ 0 \end{pmatrix}$ and one has the wanted path.

## 6  a Hierarchization in Valued Graphs

One can develop the very same algorithm in valued graph when the edges are valued with integers :

as before, we define: let $G = (\mathcal{V}, \mathcal{E})$ be a graph and let $\sim$ be an equivalence relation among the vertices of $G$. Let us consider the graph $G' = (\mathcal{E}, \mathcal{E}')$ whose vertices are the equivalence classes of $G/\sim$. Given two vertices $v_1$ and $v_2$ of $G'$, the edge $(v_1, v_2, x) \in \mathcal{E}'$ if and only if there exists $w_1 \in v_1 \subset \mathcal{V}$ and $w_2 \in v_2 \subset \mathcal{V}$ such that $(w_1, w_2) \in \mathcal{E}$ and $x$ is the minimal value of such edges.

Then one computes the paths exactly as preceedingly. The paths are refined in increasing cost, and it allows to compute the shortest path in linear time.

This can be done because when there exists a path of cost $k$ in $G$ between two vertices, then there exists a path of cost at most $k$ between the matching vertices in $G'$.

## 7  Complexity

Let $G$ be a graph consisting in $V$ vertices and let $k$ be the average number of connected edges by an outgoing edge by vertex.

Let us suppose that we built thickenings of $G$, $G_0 = G$, $G_1$, ..., by grouping the vertices two by two. The number of such graphs is $\ln(V)/ln(2)$. The number of vertices of $G_i$ is $V_i = \frac{V}{2^i}$, the average number edges of $G_i$ is $e_i$, and the average number of connected vertices by an outgoing edge is $k_i = \frac{e_i}{V_i}$.

One has $e_{i+1} = e_i - c_i$ where $c_i$ is the number of couples $v, v' \in \mathcal{V}_i$ such that $o_v \wedge o_{v'} \neq 0$.

The probability, for a given $v$ to be coupled with a vertices $v'$ such that $o_v \wedge o_{v'} \neq 0$ is $2\frac{e_i}{V_i^2} - (\frac{e_i}{V_i^2})^2$. Indeed, the probability is that $v'$ (or $v$) should be connected with of of the vertices connected to $v$ (or $v'$), but without counting twice the $v'$ connected to two different vertices connected to $v$.

Then the number $c_i$ is averagely $c_i = \frac{2e_i}{v_i^2} - \frac{e_i^2}{v_i^3}$.

Finally, $e_{i+1} = e_i - 2\frac{e_i}{V_i} + \frac{e_i^2}{V_i^3}$, and $V_i = \frac{V_O}{2^i}$.

So the probability that a path of length $l \geq 2$ of $G_i$ could be refined in two different paths of $G_{i+1}$ is less than $\left(2\frac{e_i}{v_i^2} - (\frac{e_i}{v_i^2})^2\right)^2$. This means that it may be ignored while computing the complexity function.

This also means to say that averagely, one path only will be found at each step while refining the initial path of $G_{\lfloor \ln_2(V) \rfloor}$.

So the complexity to find a path of length $l$ between two vertices is majored by $l\ln(V)$.

**Remark 4** *Let $G = G_0$ be a graph, and let $G_1, G_2, \ldots G_\infty$ be its thickenings for some equivalence relations (it is convenient to think about it as grouping the vertices two by two as in the preceding examples).*

*Let $v$ and $v'$ be two vertices between which we are looking for a path. Instead of refining paths of $G_\infty$, it is much faster to refine paths found in $G_{k_0}$ thanks to* ShortestPath1 *or* ShortestPath2. *The best way to find $k_0$ seems to be the greater integer such that the probability that two vertices are connected to the same is lower than $\frac{1}{2}$.*

## 8 Performances

These algorithms were programmed on a personal computer, having one processor athlon 2 Ghz, 512 MB or memory. The graph of $N$ vertices was built the following way : each vertex $i$ was connected to $i + 1$, $2i$, $3i$, $\ldots ki$ (all computations made modulus $N$). $k$ was chosen in several ways. This choice of graph was to complicate the computation of the paths : the multiplications commute, which increases very much the number of distinct shortest paths between two vertices. In this experimentation, the edges were all valued with 1.

The performances were :

For a graph of one million vertices and three million edges, the average time to compute the path between two vertices was 0.007 second. The number of paths was lying between 1 and 1600, for few hundreds of trials.

For a graph of ten million vertices and sixty million edges, the average time to compute the path between two vertices was 0.09 second. The number of paths was lying between 1 and 130 for a few hundreds of trials.

# 9  Conclusion

This paper has given an algorithm allowing one to compute in $D \log V$-time (with $V$ the number of edges of the graph and $D$ the diameter of the graph) the shortest path between two edges. A future work is to try to apply these ideas on flowing problems.

# 10  Akwnolegement

The author would like to thank very warmly Teresa de Diego for numerous comments on an earlier version of this paper.